\documentclass[a4paper,12pt]{article}
\usepackage{graphicx}
\usepackage{amssymb}
\usepackage{amsmath}
\usepackage{cite}
\usepackage{axodraw}
\def\be{\begin{equation}}
\def\ee{\end{equation}}
\def\bea{\begin{eqnarray}}
\def\eea{\end{eqnarray}}
\def\beb{\begin{eqnarray*}}
\def\eeb{\end{eqnarray*}}

\newcommand{\req}[1]{(\ref{#1})}
\renewcommand{\Re}{{\cal R}}
\newcommand{\newrho}{{\Re}}
\newcommand{\Bmn}{{B}}
\newcommand{\fmn}{{\cal A}}
                                                                                                                                                                                                                                                                              
\newlength{\myVSpace}
\begin{document}
\makeatletter
\def\fmslash{\@ifnextchar[{\fmsl@sh}{\fmsl@sh[0mu]}}
\def\fmsl@sh[#1]#2{%
  \mathchoice
    {\@fmsl@sh\displaystyle{#1}{#2}}%
    {\@fmsl@sh\textstyle{#1}{#2}}%
    {\@fmsl@sh\scriptstyle{#1}{#2}}%
    {\@fmsl@sh\scriptscriptstyle{#1}{#2}}}
\def\@fmsl@sh#1#2#3{\m@th\ooalign{$\hfil#1\mkern#2/\hfil$\crcr$#1#3$}}
\makeatother
\thispagestyle{empty}
\begin{titlepage}

\boldmath
\begin{center}
  {\Large {\bf The Standard Model on Non-Commutative Space-Time:
  \\[0.3cm]
  Strong Interactions Included}}
\end{center}
\unboldmath
\renewcommand{\thefootnote}{\fnsymbol{footnote}}
\begin{center}
                                                                                                                                         
{{{\bf
Bla\v zenka Meli\' c${}^1$\footnote{email: melic@thphys.irb.hr},
Kornelija Passek-Kumeri\v{c}ki${}^1$\footnote{email: passek@thphys.irb.hr},\\
Josip Trampeti\'{c}${}^{1,2}$\footnote{email:
josipt@rex.irb.hr},
Peter Schupp${}^3$\footnote{email: p.schupp@iu-bremen.de}\\ and\\
Michael\ Wohlgenannt${}^4$\footnote{email: michael.wohlgenannt@univie.ac.at}
}}}
                                                                                                                                         
\end{center}
\setcounter{footnote}{0}
\renewcommand{\thefootnote}{\arabic{footnote}}
\vskip 1em
\begin{center}
${}^{1}$Rudjer Bo\v{s}kovi\'{c} Institute, Theoretical Physics Division, \\
        P.O.Box 180, 10002 Zagreb, Croatia\\
    \vspace{.3cm}
${}^{2}$Max Planck Institut f\"{u}r Physik, \\
F\"{o}hringer Ring 6,
D-80805 M\"{u}nchen, Germany\\
    \vspace{.3cm}
${}^{3}$International University Bremen,\\ Campus Ring 8, 28759 Bremen,
    Germany\\
    \vspace{.3cm}
${}^{4}$Universit\"at Wien, Institut f\"ur theoretische Physik, \\
  Boltzmanngasse 5, 1090 Wien, Austria\\
                                                                                                                                         
\end{center}

\vspace{1cm}
                                                                                                                                         
\begin{abstract}
\noindent
This paper is a direct extension of our paper 
"The Standard Model on Non-Commutative Space-Time:
Electroweak currents and Higgs sector" \cite{Melic:2005fm}, 
now with strong interactions included.
Apart from the non-commutative corrections to Standard Model strong interactions,
several new interactions appear. The most interesting ones are gluonic interactions with the
electroweak sector. They are elaborated here in detail and the Feynman rules for 
interactions up to ${\cal O}(g^2_s \theta)$ are provided.
\end{abstract}

\end{titlepage}

\section{Introduction}
\label{sec:intro}


This paper closely follows the paper by B. Meli\'c {\it
et al}: "The Standard Model on Non-Commutative Space-Time:
Electroweak currents and Higgs sector" \cite{Melic:2005fm}, by including strong
interactions.
In this paper we
present a careful discussion of QCD and QCD-electroweak charged
and neutral currents in the Non-Com\-mu\-ta\-tive Standard Model
(NCSM) \cite{Calmet:2001na} and compute the corresponding Yukawa parts of the
action.

All relevant expressions are given in terms of
physical fields and selected Feynman rules are provided with the aim
to make the model more accessible to phenomenological
considerations.

In Sec.\ref{sec:NCSM} we briefly review the NCSM. The non-commutative QCD and 
QCD interactions with the electroweak and Yukawa sector are presented in Sections \ref{sec:ew} and 
\ref{sec:Y}. 
The selected Feynman rules to ${\cal O}(g^2_s \theta)$ 
are summarized in Sec. \ref{sec:FR}, while Sec. \ref{sec:con} is devoted to concluding remarks.

\section{Non-commutative Standard Model}
\label{sec:NCSM}
The action of the Non-Com\-mu\-ta\-tive Standard Model (NCSM) 
formally resembles the action of the
classical Standard Model (SM): the usual point-wise products in the Lagrangian are
replaced by the Moyal-Weyl product and (matter and gauge) fields are
replaced by non-commutative fermion and gauge fields (denoted by a
hat) which are expressed (via Seiberg-Witten maps) in terms of
ordinary fermion and gauge fields $\psi$ and $V_\mu$:
\begin{eqnarray}
\widehat \Psi & = & \widehat \Psi[V] = \psi - \, \frac{1}{2} \,
\theta^{\alpha \beta} \,
       V_{\alpha}  \,\partial_{\beta} \psi
          + \frac{i}{8}  \,\theta^{\alpha \beta} \,
              [V_{\alpha}, V_{\beta}]  \,\psi
+ \mathcal O(\theta^2)
\, ,\\
\widehat V_\mu & = & \widehat V_\mu[V] = V_\mu +
    \frac{1}{4} \theta^{\alpha \beta}
           \{ \partial_{\alpha}V_{\mu} + F_{\alpha \mu},V_{\beta}
           \}
+ \mathcal O(\theta^2) \, .
\end{eqnarray}
Here, $F_{\mu \nu}$ is the ordinary field strength and 
$V_\mu$ is the gauge potential corresponding to the SM gauge group structure
\begin{equation}
V_{\mu}(x)=g' {\cal A}_{\mu}(x) Y
          + g \sum_{a=1}^{3} B_{\mu}^a(x) \, T_{L}^a
          + g_s \sum_{b=1}^{8} G_{\mu}^{b}(x) \, T_S^b
\, ,\label{eq:Vmu}
\end{equation}
and $T_{L}^a=\tau_a/2$, $T_S^b=\lambda^b/2$, with $\tau^ a$ and
$\lambda^b$ being the Pauli and Gell-Mann matrices, respectively.
Here, ${\cal A}_{\mu}$, $B_{\mu}^a$ and $G_{\mu}^{b}$ represent
ordinary $U(1)_Y$, $SU(2)_L$ and $SU(3)_C$ fields, respectively.

The approach to non-commutative field theory based on star products
and Seiberg-Witten maps allows the generalization of the SM 
of particle physics to the case of  non-commutative
space-time, keeping the original gauge group and particle content
\cite{Kontsevich:1997vb,Seiberg:1999vs,Madore:2000en,Jurco:2000ja,
Jurco:2001rq,Calmet:2001na,Aschieri:2002mc,
Chaichian:2001aa,Chaichian:2004za,Chaichian:2004yw}. 
It provides a systematic way to compute Lorentz violating operators
that could be a signature of a (hypothetical) non-commutative
space-time structure
\cite{Mocioiu:2000ip,Carlson:2001sw,Anisimov:2001zc,
Armoni:2000xr,Behr:2002wx,Schupp:2002up,Minkowski:2003jg,
Trampetic:2002eb,Schupp:2004dz,
Duplancic:2003hg,Ohl:2004tn}. 
In this paper we do
not repeat the derivation, but take NCSM action
$
S_{\mbox{\tiny NCSM}}= S_{\mbox{\tiny fermions}} 
+ S_{\mbox{\tiny gauge}} + S_{\mbox{\tiny{Higgs}}} +
S_{\mbox{\tiny{Yukawa}}}
$,
given already in terms of commutative SM fields, as a starting point.
Note that the $S_{\mbox{\tiny NCSM}}$ action is anomaly free \cite{Brandt:2003fx}.

With respect to the gauge sector there are two models: the minimal
Non-Commutative Standard Model (mNCSM) \cite{Calmet:2001na} and the
non-minimal Non-Commutative Standard Model (nmNCSM)
\cite{Behr:2002wx}. The main difference between the two models is due to
the freedom of the choice of traces in the kinetic terms for gauge
fields. 
In the mNCSM, we adopt the representation that yields a model as close as 
possible to the SM without new triple gauge boson couplings.
In the nmNCSM \cite{Behr:2002wx}
the trace is chosen over all particles
on which covariant derivatives act
and which have different quantum numbers.
In the SM, these are five multiplets for each generation 
of fermions and one Higgs multiplet. 

Note that the fermion sector of the action $S_{\mbox{\tiny NCSM}}$ 
is not affected by choosing different
traces over the representations in the gauge part of the action and 
remains the same in both models. 

\subsection{Minimal/Non-Minimal NCSM}

The minimal NCSM gauge action is given by \cite{Melic:2005fm}
\bea
S^{\mbox{\tiny mNCSM}}_{\mbox{\tiny gauge}} &=&
-\frac12 \int d^4x \left( \frac12 \fmn_{\mu\nu}\fmn^{\mu\nu}
+ \, \mbox{Tr}\, {\Bmn}_{\mu\nu} {\Bmn}^{\mu\nu}
+ \, \mbox{Tr}\, G_{\mu\nu} G^{\mu\nu} \right)
\nonumber \\
&& + \frac14 \, g_s\, {d^{abc}} \, \theta^{\rho\sigma}
\int d^4x
\left(
\frac{1}{4}G^a_{\rho\sigma} G^b_{\mu\nu}
-G^a_{\rho\mu} G^b_{\sigma\nu}
\right)G^{\mu\nu,c}
+ \, {\cal O}(\theta^2) \, ,
\qquad
\label{eq:SmNCSMa}
\eea
where, $\fmn_{\mu\nu}$, ${\Bmn}_{\mu\nu}(={\Bmn}_{\mu\nu}^aT_L^a)$
and $G_{\mu\nu}(=G_{\mu\nu}^aT_S^a)$ denote field strengths.

In the non-minimal NCSM, the gauge action (\ref{eq:SmNCSMa}) is extended by new gauge terms,
Eq. (27) in \cite{Melic:2005fm}, 
from where the Lagrangians including gluons are derived 
\begin{eqnarray}
{\cal L}_{\gamma gg}&=&\frac{e}{4} \sin2{\theta_W}\,{\rm K}_{\gamma gg}\,
{\theta^{\rho\sigma}}
\left[2A^{\mu\nu}\left(2G^a_{\mu\rho}G^b_{\nu\sigma}-G^a_{\mu\nu}G^b_{\rho\sigma}\right)
\right.
\nonumber\\
& & 
+\left. 
8 A_{\mu\rho}G^{\mu\nu,a}G^b_{\nu\sigma}-A_{\rho\sigma}G^a_{\mu\nu}G^{\mu\nu,b}\right]\delta^{ab}
\, ,
\label{L12}
\\
{\cal L}_{Zgg}&=&{\cal L}_{\gamma gg}(A_{\mu}\rightarrow Z_{\mu}
)
\, ,
\label{L3456}
\end{eqnarray}
with the following coupling constants
\begin{eqnarray}
{\rm K}_{\gamma gg}&=&
\frac{-g^2_s}{2 g g'}\;
\left({g'}^2+g^2\right)\kappa_3\,,\;\;\;
{\rm K}_{Zgg}=-\frac{g'}{g}\;{\rm K}_{\gamma gg}
\, .
\label{K123456}
\end{eqnarray}
Details of the derivations of neutral triple-gauge boson terms and
the properties of the coupling constants
in (\ref{K123456}) are given in \cite{Behr:2002wx,Duplancic:2003hg}.


\section{QCD and QCD-Electroweak Matter Currents}
\label{sec:ew}

The general fermionic action in NCSM reads \cite{Melic:2005fm}
\begin{eqnarray}
S_{\mbox{\tiny $\psi$}}
=\int d^4x \left( i \overline{\psi} \, \fmslash D \, \psi
-\frac{i}{4}\,  \overline{\psi}\,
 \theta^{\mu \nu \rho} \, {\newrho}_{\psi}(F_{\mu \nu}) \, D_{\rho} \psi \right)
+ {\cal O}(\theta^2)
\, ,
\label{eq:hatPsiDPsi}
\end{eqnarray}
where
$\theta^{\mu \nu \rho}=
\theta^{\mu \nu} \gamma^{\rho}
+ \theta^{\nu \rho} \gamma^{\mu}
+ \theta^{\rho \mu} \gamma^{\nu}$
is a totally antisymmetric quantity,
$\psi$ denotes any fermion field,
while the corresponding representations
$\newrho_{\psi}$ for various fermion fields are listed
in Table 2 of Ref. \cite{Melic:2005fm}.

We express the NCSM currents
in terms of physical fields
starting with the left-handed sector.
In (\ref{eq:hatPsiDPsi}),
the representation ${\newrho}_{\Psi_L}(V_{\mu})$ of the SM
gauge group takes the form
\begin{equation}
{\newrho}_{\Psi_L}(V_{\mu})= g' \, {\cal A}_{\mu} \, Y_{\Psi_L} +
g\, B_{\mu}^a T_L^{a} + g_s\, G_{\mu}^b T_S^{b}\, .\label{eq:left}
\end{equation}
The hypercharge generator $Y_{\Psi_L}$
can be rewritten as
$
Y_{\Psi_L}=Q_{q_{\mbox{\tiny up}}}-T_{3,q_{\mbox{\tiny up},L}}
        =Q_{q_{\mbox{\tiny down}}}-T_{3,q_{\mbox{\tiny down},L}}$.
Then the left-handed electroweak part
of the action $S_{\mbox{\tiny $\psi$}}$
can be cast in the form
\begin{eqnarray}
S_{\mbox{\tiny $\psi$,ew,L}}
 &=&
 \int d^4 x \left( \bar\Psi_L\, i\fmslash \partial\, \Psi_L +
\bar\Psi_L \, {\mathbf J}^{(L)} \, \Psi_L \right)\,,
\nonumber \\
{\mathbf J}^{(L)} &=&
\bar q_{\mbox{\tiny up},L} \, J_{12}^{(L)} \, q_{\mbox{\tiny down},L}
+
\bar q_{\mbox{\tiny down},L} \, J_{21}^{(L)} \, q_{\mbox{\tiny up},L}
\nonumber \\
 &&
\, + \,
\bar q_{\mbox{\tiny up},L} \, J_{11}^{(L)} \, q_{\mbox{\tiny up},L}
+
\bar q_{\mbox{\tiny down},L} \, J_{22}^{(L)} \, q_{\mbox{\tiny down},L}
\, ,
\label{eq:SewL}
\end{eqnarray}
where ${\mathbf J^{(L)}}$
is a $2\times 2$ matrix whose
off-diagonal elements ($J_{12}^{(L)}$, $J_{21}^{(L)}$)
denote
the charged currents
and diagonal elements ($J_{11}^{(L)}$, $J_{22}^{(L)}$)
the neutral currents.
After some algebra, with gluons included ($G_{\mu}=G_{\mu}^a T_S^a$), we obtain

\begin{subequations}
\begin{eqnarray}
J_{12}^{(L)} &=& \frac{g}{\sqrt{2}} \fmslash W^+
           + J_{12}^{(L,{\theta})}
           + {\cal O}(\theta^2)
\, ,
\label{eq:JL12}
\\[0.2cm]
J_{21}^{(L)} &=& \frac{g}{\sqrt{2}} \fmslash W^-
           + J_{21}^{(L, {\theta})}
           + {\cal O}(\theta^2)
\, ,
\label{eq:JL21}
\end{eqnarray}
\end{subequations}
\begin{subequations}
\begin{eqnarray}
J_{11}^{(L)} &=& \left[
           e \, Q_{q_{\mbox{\tiny up}}} \, \fmslash A
           + \frac{g}{\cos \theta_W}
           (T_{3,q_{\mbox{\tiny up},L}}
           -  Q_{q_{\mbox{\tiny up}}}\,  \sin^2 \theta_W)
            {\fmslash Z} + g_s {\fmslash G}
           \right]
\nonumber \\ & &
           + \,J_{11}^{(L,{\theta})}
 + {\cal O}(\theta^2),
\label{eq:JL11}
       \\[0.2cm]
\nonumber
J_{22}^{(L)} &=& \left[
           e \, Q_{q_{\mbox{\tiny down}}} \, \fmslash A
           + \frac{g}{\cos \theta_W}
           (T_{3,q_{\mbox{\tiny down},L}}
           -  Q_{q_{\mbox{\tiny down}}}\,  \sin^2 \theta_W)
            {\fmslash Z} + g_s {\fmslash G}
           \right]
\nonumber \\ & &
           + \,J_{22}^{(L,{\theta})}
           + {\cal O}(\theta^2),
\label{eq:JL22}
\end{eqnarray}
\label{eq:JLcurr}
\end{subequations}
where $J_{12}^{(L,{\theta})}$ and $J_{11}^{(L,{\theta})}$, 
Eqs. (41-43) in \cite{Melic:2005fm}, receive the following additional
contributions from the inclusion of gluons  
\begin{eqnarray}
J_{12}^{(L,{\theta})} & : &\;\;\; \frac{g}{2\sqrt{2}} \, \theta^{\mu \nu \rho}
   \, W_{\mu}^+
 \left\{
 g_s \left[ G_{\nu}  \;
 (\stackrel{\leftarrow}{\partial}_{\rho} +
 \stackrel{\rightarrow}{\partial}_{\rho}) + 2({\partial}_{\rho} G_{\nu})
  \right]
  \right.
\nonumber
  \\ & &
  \left.
 - i\,e\,g_s (Q_{q_{\mbox{\tiny up}}}
- Q_{q_{\mbox{\tiny down}}}) A_{\nu}G_{\rho}
  \right.
\nonumber
  \\ & &
  \left.
- i\,g_s\, \frac{g}{\cos \theta_W} \,
  \bigg(
           (T_{3,q_{\mbox{\tiny up},L}}
- T_{3,q_{\mbox{\tiny down},L}})
           -  (Q_{q_{\mbox{\tiny up}}}
- Q_{q_{\mbox{\tiny down}}})
\,  \sin^2 \theta_W
       \bigg)
    \;  Z_{\nu}G_{\rho}
    \right.
\nonumber
  \\ & &
  \left.
  + i\,g^2_s\, G_{\nu}G_{\rho}
   \right\}\,,
\label{eq:JL12theta}
\\
\nonumber
\\
J_{11}^{(L,{\theta})} & : &\;\;\; \frac{1}{2} \, \theta^{\mu \nu \rho}
 \left\{
  i\, g_s \: (\partial_{\nu} G_{\mu}) \,
   \stackrel{\rightarrow}{\partial}_{\rho}
  \right.
\nonumber
  \\ & &
  \left.
 - g^2_s\,\left[G_{\mu} G_{\nu} \,
   \stackrel{\rightarrow}{\partial}_{\rho}
   +({\partial}_{\rho}G_{\mu}) G_{\nu}\right]
   + ig^3_s\,G_{\mu} G_{\nu}G_{\rho}
  \right.
\nonumber
  \\ & &
  \left.
 - \, e \,g_s\, Q_{q_{\mbox{\tiny up}}}
      \: \left[(\partial_{\rho} A_{\mu}) \, G_{\nu}
      -A_{\mu} \,(\partial_{\rho} G_{\nu})\right]
  \right.
\nonumber
  \\ & &
  \left.
 - \, \frac{g}{\cos \theta_W} \,g_s
           (T_{3,q_{\mbox{\tiny up},L}}
           -  Q_{q_{\mbox{\tiny up}}}\,  \sin^2 \theta_W)
      \: \left[(\partial_{\rho} Z_{\mu}) \, G_{\nu}
      -Z_{\mu}(\partial_{\rho} \, G_{\nu})\right]
  \right.
\nonumber
  \\ & &
  \left.
 + i\,\frac{g^2}{2}\,g_s\,\; W_{\mu}^+ \, W_{\nu}^- \, G_{\rho}
 +i\,e\,Q_{q_{\mbox{\tiny up}}} \,g^2_s \,A_{\mu} \, G_{\nu} \, G_{\rho}
   \right.
\nonumber
  \\ & &
  \left.
 + i\, \frac{g}{\cos \theta_W} \,g_s^2
           (T_{3,q_{\mbox{\tiny up},L}}
           -  Q_{q_{\mbox{\tiny up}}}\,  \sin^2 \theta_W)
      \:  Z_{\mu} \, G_{\nu}\, G_{\rho}
   \right\}
\, ,
\label{eq:JL11theta}
\end{eqnarray}
while
\begin{equation}
\left.
\begin{array}{l}
J_{21}^{(L,{\theta})} \\[0.2cm]
J_{22}^{(L,{\theta})}
\end{array}
\right\}
 =
\left\{
\begin{array}{l}
J_{12}^{(L,{\theta})} \\[0.2cm]
J_{11}^{(L,{\theta})}
\end{array}
\right.
(
W^+  \leftrightarrow  W^- ,
 Q_{q_{\mbox{\tiny up}}}
 \leftrightarrow  Q_{q_{\mbox{\tiny down}}} ,
 T_{3,q_{\mbox{\tiny up},L}}
 \leftrightarrow  T_{3,q_{\mbox{\tiny down},L}}
)\, .
\label{eq:JL21thetaJL22theta}
\end{equation}
In this paper we use the notation
$({\partial}_{\rho} q \equiv
\stackrel{\rightarrow}{\partial}_{\rho} q)$ and $({\partial}_{\rho} \overline{q} \equiv
\overline{q} \stackrel{\leftarrow}{\partial}_{\rho})$.

For the right-handed sector $q_R$ represents 
$q_R \in \{q_{\mbox{\tiny up},R},q_{\mbox{\tiny down},R} \}$,
and the representation ${\newrho}_{q_R}(V_{\mu})$ is given by
\begin{equation}
{\newrho}_{q_R}(V_{\mu})=
g' \, {\cal A}_{\mu}  +  Y_{q_R}\,g_s\, G_{\mu}
\, , 
\label{eq:right}
\end{equation}
with SU(3) fields included.
Note that for the right-handed fermions we have
$T_{3,q_R}=0$ and $Y_{q_R}=Q_{q}$.
The right-handed electroweak part
of the action $S_{\mbox{\tiny $\psi$}}$
\begin{eqnarray}
S_{\mbox{\tiny $\psi$,ew,R}}
 &=&
 \int d^4 x \left( \bar q_R\, i\fmslash \partial\, q_R +
\bar q_R \, J^{(R)} \, q_R \right)\,,
\nonumber\\
J^{(R)} &=& \left[
           e \, Q_{q} \, \fmslash A
           -eQ_{q} \, \tan\theta_W
            {\fmslash Z} + g_s\,{\fmslash G}
        \right]
           + J^{(R,{\theta})}
           + {\cal O}(\theta^2)
\, ,
\label{eq:JRcurr}
\end{eqnarray}
has the following additional gluon contributions to Eq. (48) in \cite{Melic:2005fm} 
\begin{eqnarray}
J^{(R,{\theta})} & : &\;\;\; \frac{1}{2} \, \theta^{\mu \nu \rho}
 \left\{
  i\, g_s \: (\partial_{\nu} G_{\mu}) \,
   \stackrel{\rightarrow}{\partial}_{\rho}
  \right.
\nonumber
  \\ & &
  \left.
 - g^2_s\,\left[G_{\mu} G_{\nu} \,
   \stackrel{\rightarrow}{\partial}_{\rho}
   +({\partial}_{\rho}G_{\mu}) G_{\nu}\right]
   + ig^3_s\,G_{\mu} G_{\nu}G_{\rho}
  \right.
\nonumber
  \\ & &
  \left.
 - \,g_s\,e\,Q_{q}
      \: \left[(\partial_{\rho} A_{\mu}) \, G_{\nu}
      -A_{\mu} \,(\partial_{\rho} G_{\nu})\right]
  \right.
\nonumber
  \\ & &
  \left.
 +
 \,g_s
       \,e \,Q_{q}\,\tan\theta_W
      \: \left[(\partial_{\rho} Z_{\mu}) \, G_{\nu}
      -Z_{\mu}(\partial_{\rho} \, G_{\nu})\right]
  \right.
\nonumber
  \\ & &
  \left.
 +i\,g^2_s \,e\,Q_{q} \,\left[A_{\mu} -\tan\theta_W Z_{\mu}\right]\, G_{\nu} \, G_{\rho}
   \right\}
\, .
\label{eq:JRtheta}
\end{eqnarray}
\section{Yukawa Terms}
\label{sec:Y}

By the similar analysis as in the preceeding section, one can show
that Yukawa part of the action 
\begin{eqnarray}
S_{\mbox{\tiny $\psi$, Yukawa}} & = &
\int d^4 x \sum_{i,j=1}^3
\left [
\bar d^{(i)}
\left(
N_{dd}^{V(ij)}
+ \gamma_5 \,
N_{dd}^{A(ij)}
\right)
d^{(j)}
\right.
\nonumber \\[0.1cm] & & \left.
+ \bar u^{(i)}
\left(
N_{uu}^{V(ij)}
+ \gamma_5 \,
N_{uu}^{A(ij)}
\right)
u^{(j)}
\right.
\nonumber \\[0.1cm] & & \left.
+ \bar u^{(i)}
\left(
C_{ud}^{V(ij)}
+ \gamma_5 \,
C_{ud}^{A(ij)}
\right)
d^{(j)}
\right.
\nonumber \\[0.1cm] & & \left.
+ d^{(i)}
\left(
C_{du}^{V(ij)}
+ \gamma_5 \,
C_{du}^{A(ij)}
\right)
u^{(j)}
\right]\, ,
\label{eq:SYukawa-fiz}
\end{eqnarray}
contains additional gluon contributions to the neutral currents
(Eqs. (67-69) from \cite{Melic:2005fm}):
\begin{eqnarray}
{N_{dd}^{V,{\theta} (ij)}}
&:&
- \frac{1}{2} \,g_s\,\theta^{\mu \nu}
M_{\mbox{\tiny down}}^{(ij)}
\left\{
 -\,G_{\mu}\,\frac{(\partial_{\nu}h)}{v}
 +\left[\partial_{\nu}\,G_{\mu}\,+\,ig_s\,G_{\mu}\,\,G_{\nu}
 \right]\left(1+\frac{h}{v}\right)
\right\}
\, ,
\nonumber
\\
\label{eq:NdV}
\\
N_{dd}^{A,{\theta} (ij)}
&:&
\frac{igg_s}{2 \cos \theta_W}
\,
 \theta^{\mu \nu}
M_{\mbox{\tiny down}}^{(ij)}
\;
T_{3,\psi_{\mbox{\tiny down},L}}
\,
Z_{\mu}G_{\nu}
\left( 1 + \frac{h}{v}\right)
\,,
\label{eq:NuVA}
\end{eqnarray}
as well as to the charge currents, Eq. (71,72) in \cite{Melic:2005fm}
\begin{eqnarray}
C_{ud}^{V,{\theta} (ij)}&:&\;
- \frac{i g g_s}{4 \sqrt{2}}
\theta^{\mu \nu}
\left( 1 + \frac{h}{v}\right)
\left[
(V M_{\mbox{\tiny down}})^{(ij)}
-
(M_{\mbox{\tiny up}} V)^{(ij)}
\right]
G_{\mu} W_{\nu}^+
\,,\nonumber\\
C_{ud}^{A,{\theta} (ij)} &:&\;
- \frac{i g g_s}{4 \sqrt{2}}
\theta^{\mu \nu}
\left( 1 + \frac{h}{v}\right)
\left[
(V M_{\mbox{\tiny down}})^{(ij)}
+
(M_{\mbox{\tiny up}} V)^{(ij)}
\right]
G_{\mu} W_{\nu}^+
\, .
\nonumber \\
\end{eqnarray}
In the above, $M_{\mbox{\tiny up,down}}^{(ij)}$ are usual $3\times 3$ diagonal
mass matrices defined in \cite{Melic:2005fm}, and 
\begin{eqnarray}
\left.
\begin{array}{l}
 N_{uu}^{V,{\theta} (ij)}\\[0.2cm]
N_{uu}^{A,{\theta} (ij)}
\end{array}
\right\}
& =&
\left\{
\begin{array}{l}
N_{dd}^{V,{\theta} (ij)}\\[0.2cm]
N_{dd}^{A,{\theta} (ij)}
\end{array}
\right.
(
W^+  \leftrightarrow  W^-,
{\mbox{down}}
\rightarrow
\mbox{up}
 ),
 \nonumber\\
C_{du}^{V (ij)} & = &
\left( C_{ud}^{V (ij)}
(\stackrel{\rightarrow}{\partial}
\leftrightarrow
\stackrel{\leftarrow}{\partial}
)\right) ^{\dagger}
\, ,
\nonumber \\
C_{du}^{A (ij)} & = & -
\left( C_{ud}^{A (ij)}
(\stackrel{\rightarrow}{\partial}
\leftrightarrow
\stackrel{\leftarrow}{\partial}
)\right) ^{\dagger}\, ,
\label{eq:ChVA}
\end{eqnarray}
while $V^{(ij)}$ are the standard CKM matrix elements.

\section{Feynman Rules}
\label{sec:FR}

In this section, we list
a number of selected Feynman rules for non-commutative QCD
and non-commutative QCD-electroweak sectors
up to the first order in
$\theta$ and up to $g^2_s$ order.
Higher-order terms are not
considered in this work. 

The following notation for vertices has been adopted:
{\it all gauge boson lines are taken to be incoming;
the momenta of the incoming and outgoing quarks,
following the flow of the quark line
are given by $p_{\mbox{\tiny {\rm in}}}$ and
$p_{\mbox{\tiny {\rm out}}}$, respectively}.
In the following we denote quarks by $q \in \{ u^{(i)}, d^{(i)} \}$,
and the generation indices by $i$ and $j$.
In the Feynman rules we use the following definitions 
\begin{eqnarray}
c_{V,q} & = &
T_{3,q_{L}}\, -\,  2\,  Q_{q}\,  \sin^2 \theta_W\,,
\nonumber\\
c_{A,q} & = & T_{3,q_L}
\,.
\label{cva}
\end{eqnarray}
We also make use of 
$(\theta k)^\mu\equiv\theta^{\mu\nu}k_\nu 
=- k_\nu\theta^{\nu\mu}\equiv -(k\theta)^\mu$,
$(k \theta p)\equiv k_\mu \theta^{\mu\nu} p_{\nu}$ and 
$\theta^{\mu \nu \rho}=
\theta^{\mu \nu} \gamma^{\rho}
+ \theta^{\nu \rho} \gamma^{\mu}
+ \theta^{\rho \mu} \gamma^{\nu}$.


\subsection{Minimal NCSM}

The $\theta$ corrections to vertices containing quarks
are obtained using Eqs.  (\ref{eq:SewL},\ref{eq:JRcurr}). The
Yukawa part  of the action \req{eq:SYukawa-fiz} has to be taken into
account as well, because it generates additional mass proportional
terms which modify some interaction vertices. In comparison with
the SM, this is a novel feature. 
According to \req{eq:SmNCSMa},
the gauge boson couplings
receive $\theta$
proportional corrections to three- and four-gluon couplings. 

First, let us define the three-gauge boson vertex function
\begin{eqnarray}
\Theta_3((\mu,k_1),(\nu,k_2),(\rho,k_3)) &=&
\nonumber \\
& & \hspace*{-4cm}
-\,(k_1 \theta k_2)\,
[(k_1-k_2)^\rho g^{\mu \nu} +(k_2-k_3)^\mu g^{\nu \rho} + (k_3-k_1)^\nu g^{\rho \mu}]
\nonumber \\
& & \hspace*{-4cm}
-\,\theta^{\mu \nu}\,
[ k_1^\rho \, (k_2 k_3) - k_2^\rho \, (k_1 k_3) ]
\nonumber \\
& & \hspace*{-4cm}
-\,\theta^{\nu \rho}\,
[ k_2^\mu \, (k_3 k_1) - k_3^\mu \, (k_2 k_1) ]
\nonumber \\
& & \hspace*{-4cm}
-\,\theta^{\rho \mu}\,
[ k_3^\nu \, (k_1 k_2) - k_1^\nu \, (k_3 k_2) ]
\nonumber \\
& & \hspace*{-4cm}
+\,(\theta k_2)^\mu \,\left[g^{\nu \rho}\, k_3^2 - k_3^\nu k_3^\rho\right]
+(\theta k_3)^\mu\,\left[g^{\nu \rho}\, k_2^2 - k_2^\nu k_2^\rho\right]
\nonumber \\
& & \hspace*{-4cm}
+\,(\theta k_3)^\nu \,\left[g^{\mu \rho}\, k_1^2 - k_1^\mu k_1^\rho \right]
+(\theta k_1)^\nu \,\left[g^{\mu \rho}\, k_3^2 - k_3^\mu k_3^\rho \right]
\nonumber \\
& & \hspace*{-4cm}
+\,(\theta k_1)^\rho \,\left[g^{\mu \nu}\, k_2^2 - k_2^\mu k_2^\nu \right]
+(\theta k_2)^\rho \,\left[g^{\mu \nu}\, k_1^2 - k_1^\mu k_1^\nu \right]
\, ,
\label{eq:Gamma}
\end{eqnarray}
which is the same function as in
\cite{Melic:2005fm}, but written in the explicit form. 

Similarly, the result of the four-gauge boson vertex function is
\begin{eqnarray}
\Theta_4( (\mu,k_1), (\nu,k_2), (\rho,k_3), (\sigma, k_4) ) &=& 
\nonumber \\
&& \hspace*{-4cm}
(k_3\theta k_4)
\left(g^{\mu\rho}g^{\nu\sigma} - g^{\mu\sigma}g^{\nu\rho}\right)
\nonumber
\\
&& \hspace*{-4cm}
\hspace{-.2cm}
+\,\theta^{\mu\nu}
[k_3^{\sigma}k_4^{\rho} - g^{\rho\sigma} (k_3 k_4)]
+\theta^{\rho\sigma}
(k_3^{\mu} k_4^{\nu} - k_3^{\nu} k_4^{\mu})
\nonumber
\\
&& \hspace*{-4cm}
\hspace{-.2cm}
-\,\theta^{\mu\rho}
[k_3^{\sigma}k_4^{\nu} - g^{\nu\sigma} (k_3 k_4)]
-\theta^{\mu\sigma}
[k_3^{\nu}k_4^{\rho} - g^{\nu\rho} (k_3 k_4)]
\nonumber
\\
&& \hspace*{-4cm}
\hspace{-.2cm}
+\,\theta^{\nu\rho}
[k_3^{\sigma}k_4^{\mu} - g^{\mu\sigma} (k_3 k_4)]
+\theta^{\nu\sigma}
[k_3^{\mu}k_4^{\rho} - g^{\mu\rho} (k_3 k_4)]
\nonumber
\\
&& \hspace*{-4cm}
\hspace{-.2cm}
+\,(\theta k_3)^\mu \,\left(k_4^{\nu}\,g^{\rho\sigma} - k_4^{\rho}\,g^{\nu\sigma}\right)
+\,(\theta k_4)^\mu\,\left(k_3^{\nu}\,g^{\rho\sigma} - k_3^{\sigma}\,g^{\nu\rho}\right)
\nonumber
\\
&& \hspace*{-4cm}
\hspace{-.2cm}
-\,(\theta k_3)^\nu\,\left(k_4^{\mu}\,g^{\rho\sigma} - k_4^{\rho}\,g^{\mu\sigma}\right)
-\,(\theta k_4)^\nu\,\left(k_3^{\mu}\,g^{\rho\sigma} - k_3^{\sigma}\,g^{\mu\rho}\right)
\nonumber
\\
&& \hspace*{-4cm}
\hspace{-.2cm}
+\,(\theta k_3)^\rho \,\left(k_4^{\mu}\,g^{\nu\sigma} - k_4^{\nu}\,g^{\mu\sigma}\right)
-\,(\theta k_4)^\rho \,\left(k_3^{\mu}\,g^{\nu\sigma} - k_3^{\nu}\,g^{\mu\rho}\right)
\nonumber
\\
&& \hspace*{-4cm}
\hspace{-.2cm}
-\,(\theta k_3)^\sigma \,\left(k_4^{\mu}\,g^{\nu\rho} - k_4^{\nu}\,g^{\mu\rho}\right)
+\,(\theta k_4)^\sigma \,\left(k_3^{\mu}\,g^{\nu\rho} - k_3^{\nu}\,g^{\mu\rho}\right)\,.
\nonumber\\
\label{eq:fourG}
\end{eqnarray}
\newpage
The three vertices that appear in the commutative SM as well are
\noindent
\begin{itemize}
\item \hspace{-1cm}\\
\begin{picture}(55,45) (30,-30)
\SetWidth{0.5}
\ArrowLine(30,-30)(50,-10)
\ArrowLine(50,-10)(30,11)
\Gluon(80,-10)(50,-10){2}{5}
\Vertex(50,-10){1.5}
\Text(40,11)[lb]{$q$}
\Text(40,-40)[lb]{$q$}
\Text(75,-5)[lb]{$G_{\mu}^a(k)$}
\end{picture}
\begin{eqnarray}
\lefteqn{i\, g_s \,
  \left[
 \gamma_{\mu}
 - \frac{i}{2} \, k^{\nu}
   \left( \theta_{\mu \nu \rho}
 \, p_{\mbox{\tiny in}}^{\rho}-
\theta_{\mu \nu}\, m_q\,\right)
  \right]\,T^a_S}
\nonumber \\[0.2cm]
&=& 
 i \,g_s \,
\gamma_{\mu}\,T^a_S
\nonumber \\
&& 
 + \frac{1}{2} \,g_s \,
\left[
(p_{\mbox{\tiny out}} \theta p_{\mbox{\tiny in}}) \gamma_\mu
-
(p_{\mbox{\tiny out}} \theta)_\mu
(\slash \! \! \! p_{\mbox{\tiny in}}-m_q)
-
(\slash \!\!\! p_{\mbox{\tiny out}}-m_q)
(\theta p_{\mbox{\tiny in}})_\mu
\right]\,T^a_S \,,
\nonumber
\\
\label{eq:ffgluon}
\end{eqnarray}
\item \hspace{-1cm}\\
\begin{picture}(55,45) (30,-30)
\SetWidth{0.5}
\Gluon(30,-30)(50,-10){2}{5}
\Gluon(50,-10)(30,11){2}{5}
\Gluon(80,-10)(50,-10){2}{5}
\Vertex(50,-10){1.5}
\Text(40,11)[lb]{$G^c_{\rho}(k_3)$}
\Text(40,-40)[lb]{$G^a_{\mu}(k_1)$}
\Text(75,-5)[lb]{$G^b_{\nu}(k_2)$}
\end{picture}
\begin{eqnarray}
&& g_s\,f^{abc}\left[g_{\mu\nu}(k_1-k_2)_{\rho}+g_{\nu\rho}(k_2-k_3)_{\mu}
+g_{\rho\mu}(k_3-k_1)_{\nu}\right]
\nonumber
\\
&&+\frac{1}{2}\,g_s\,d^{abc}  \,
\Theta_3((\mu,k_1),(\nu,k_2),(\rho,k_3))\,.
\label{eq:3G}
\end{eqnarray}
\end{itemize}
The $\theta$-corrected $gggg$ vertex takes the following form:

\begin{itemize}
\item \hspace{4cm}\\ \\
\begin{picture}(55,45) (-20,-30)
\SetWidth{0.5}
\Gluon(30,-30)(50,-10){2}{5}
\Gluon(50,-10)(30,11){2}{5}
\Gluon(70,-30)(50,-10){2}{5}
\Gluon(70,11)(50,-10){2}{5}
\Vertex(50,-10){1.5}
\Text(-10,11)[lb]{$G^c_{\rho}(k_3)$}
\Text(-10,-40)[lb]{$G^d_{\sigma}(k_4)$}
\Text(77,-40)[lb]{$G^a_{\mu}(k_1)$}
\Text(77,8)[lb]{$G^b_{\nu}(k_2)$}
\end{picture}
\\
\begin{eqnarray}
&&
\hspace{-1cm}
i g_s^2 \left\{f^{abx}f^{cdx}\left(g^{\mu\sigma}g^{\nu\rho}-g^{\mu\rho}g^{\nu\sigma}\right)+
f^{acx}f^{bdx}\left(g^{\mu\sigma}g^{\nu\rho}-g^{\mu\nu}g^{\rho\sigma}\right)
\right.
\nonumber
\\
&&
\left.
\hspace{-.3cm}+f^{adx}f^{bcx}\left(g^{\mu\rho}g^{\nu\sigma}-g^{\mu\nu}g^{\rho\sigma}\right)
\right\}
\nonumber
\\
&&
\hspace{-1.3cm}+\,\frac{i}{2}\,g_s^2\,\left\{
f^{abx}d^{cdx}\Theta_4((\mu,k_1),(\nu,k_2),(\rho,k_3),(\sigma,k_4) )
\right.
\nonumber\\
&&
\left.
\hspace{.2cm}
+ [(\mu,k_1,a) \leftrightarrow (\rho,k_3,c)] + [(\mu,k_1,a)\leftrightarrow (\sigma,k_4,d)]
\right.
\nonumber\\
&&
\left.
\hspace{.2cm}
+  [(\nu,k_2,b) \leftrightarrow (\rho,k_3,c)] + [(\nu,k_2,b) \leftrightarrow (\sigma,k_4,d)]
\right.
\nonumber\\
&&
\left. 
\hspace{.2cm}
+ [(\mu,k_1,a) \leftrightarrow  (\rho,k_3,c) , (\nu,k_2,b) \leftrightarrow (\sigma,k_4,d)]
\right \}\,.
\label{eq:4G}
\end{eqnarray}

\end{itemize}

Equations (\ref{eq:SewL},\ref{eq:JRcurr})
describes the interaction vertices involving quarks and
two or three gauge bosons. These do not appear in the SM.
In the following we provide all
contributions to such vertices with four legs
and the corresponding mass-proportional contributions.
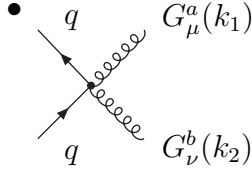
\begin{itemize}
\item \hspace{1cm}\\
\begin{picture}(55,45) (30,-30)
\SetWidth{0.5}
\ArrowLine(30,-30)(50,-10)
\ArrowLine(50,-10)(30,11)
\Gluon(70,-30)(50,-10){2}{5}
\Gluon(70,11)(50,-10){2}{5}
\Vertex(50,-10){1.5}
\Text(40,11)[lb]{$q$}
\Text(40,-40)[lb]{$q$}
\Text(77,-40)[lb]{$G^b_{\nu}(k_2)$}
\Text(77,8)[lb]{$G^a_{\mu}(k_1)$}
\end{picture}

\begin{eqnarray}
-\frac{g^2_s}{2}\,\left\{
    \theta_{\mu\nu\rho} \ (k_1^{\rho}-k_2^{\rho})\,T^a_S T^b_S
    +i\left[
    \theta_{\mu\nu\rho} \ (p_{\mbox{\tiny in}}^{\rho}+k_2^{\rho})
    -\theta_{\mu\nu}m_q\right]\,f^{abc} T^c_S \right\},
\label{eq:ffGG}
\end{eqnarray}\\

\item \hspace{1cm}\\
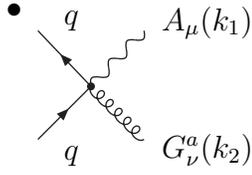
\begin{picture}(55,45) (30,-30)
\SetWidth{0.5}
\ArrowLine(30,-30)(50,-10)
\ArrowLine(50,-10)(30,11)
\Gluon(70,-30)(50,-10){2}{5}
\Photon(70,11)(50,-10){2}{3}
\Vertex(50,-10){1.5}
\Text(40,11)[lb]{$q$}
\Text(40,-40)[lb]{$q$}
\Text(77,-40)[lb]{$G^a_{\nu}(k_2)$}
\Text(77,8)[lb]{$A_{\mu}(k_1)$}
\end{picture}
\vspace{-1.5cm}
\begin{eqnarray}
-\,\frac{1}{2}\,g_s\,e\, Q_q \,
    \theta_{\mu\nu\rho} \, (k_1^{\rho}-k_2^{\rho})\,T^a_S,
\label{eq:ffgluong}
\end{eqnarray}\\

\item \hspace{1cm}\\
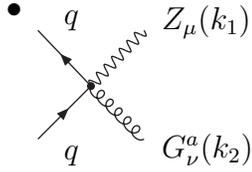
\begin{picture}(55,45) (30,-30)
\SetWidth{0.5}
\ArrowLine(30,-30)(50,-10)
\ArrowLine(50,-10)(30,11)
\Gluon(70,-30)(50,-10){2}{5}
\Photon(70,11)(50,-10){2}{7}
\Vertex(50,-10){1.5}
\Text(40,11)[lb]{$q$}
\Text(40,-40)[lb]{$q$}
\Text(77,-40)[lb]{$G^a_{\nu}(k_2)$}
\Text(77,8)[lb]{$Z_{\mu}(k_1)$}
\end{picture}

\begin{equation}
\hspace{2cm}
\frac{-\,e\,g_s }{2\sin2\theta_W}
\Big[ \theta_{\mu\nu\rho} \, (k_1^{\rho}-k_2^{\rho})(c_{V,q}-c_{A,q}\gamma_5)
+2\theta_{\mu\nu}\,m_q\,c_{A,q}\gamma_5\Big] \,T^a_S\,,
\label{eq:ffgluonZ}
\end{equation}
\\

\item\hspace{1cm}\\
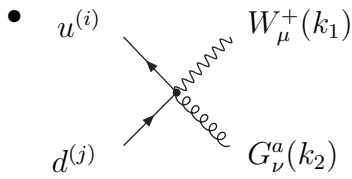
\hspace*{1cm}
\begin{picture}(55,45) (30,-30)
\SetWidth{0.5}
\ArrowLine(30,-30)(50,-10)
\ArrowLine(50,-10)(30,11)
\Gluon(70,-30)(50,-10){2}{5}
\Photon(70,11)(50,-10){2}{7}
\Vertex(50,-10){1.5}
\Text(5,11)[lb]{${u}^{(i)}$}
\Text(3,-40)[lb]{${d}^{(j)}$}
\Text(77,-40)[lb]{$G^a_{\nu}(k_2)$}
\Text(77,8)[lb]{$W^+_{\mu}(k_1)$}
\end{picture}
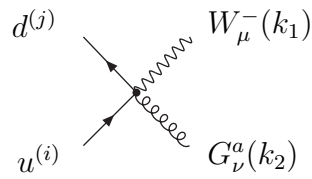
\hspace*{4cm}
\begin{picture}(55,45) (30,-30)
\SetWidth{0.5}
\ArrowLine(30,-30)(50,-10)
\ArrowLine(50,-10)(30,11)
\Gluon(70,-30)(50,-10){2}{5}
\Photon(70,11)(50,-10){2}{7}
\Vertex(50,-10){1.5}
\Text(3,11)[lb]{${d}^{(j)}$}
\Text(5,-40)[lb]{${u}^{(i)}$}
\Text(77,-40)[lb]{$G^a_{\nu}(k_2)$}
\Text(77,8)[lb]{$W^-_{\mu}(k_1)$}
\end{picture}

\begin{eqnarray}
&&
\hspace{-1cm}
\frac{- e\,g_s}{4 \sqrt{2} \sin\theta_W}
\left(
V_{ij} \atop 
V^{*}_{ij}
\right )
\left\{{{{}^{}_{}}\atop{{}^{}_{}}} \theta_{\mu \nu\rho}
(k_1^{\rho}-k_2^{\rho}) \,
 \left( 1 - \gamma_5 \right)
 \right.
\nonumber
\\
&&
\hspace{-1.cm}
\left.
-\; \theta_{\mu \nu}
\left[
{
m_{{u}^{(i)}}
 \choose
m_{{d}{^{(j)}}}
}
(1-\gamma_5)
-
{
m_{{d}^{(j)}}
  \choose
m_{{u}^{(i)}}}
  (1+\gamma_5)
\right]
 \right\}
\,T^a_S\,.
\label{eq:qqgluonW}
\end{eqnarray}

\end{itemize}

Similarly, five-field vertices
$qqWWg$, $qq\gamma gg$, $qqZgg$, $qqW\gamma g$, $qqWZg$ and $qqWgg$
are extracted from Eqs. (\ref{eq:SewL},\ref{eq:JRcurr}) as well.
They have no mass-proportional corrections:

\vspace{.5cm}
\begin{itemize}

\item \hspace{1cm}\\
\begin{picture}(55,45) (30,-30)
\SetWidth{0.5}
\ArrowLine(30,-30)(50,-10)
\ArrowLine(50,-10)(30,11)
\Photon(70,11)(50,-10){2}{7}
\Gluon(70,-30)(50,-10){2}{5}
\Photon(75,-10)(50,-10){2}{7}
\Vertex(50,-10){1.5}
\Text(40,11)[lb]{$q$}
\Text(40,-40)[lb]{$q$}
\Text(80,-15)[lb]{$W^-_{\nu}(k_2)$}
\Text(77,-40)[lb]{$G^a_{\rho}(k_3)$}
\Text(77,8)[lb]{$W^+_{\mu}(k_1)$}
\end{picture}
\vspace{-1.5cm}
\begin{equation}
\hspace{2cm}
-\frac{e^2\,g_s }{8\sin^2\theta_W}
\theta_{\mu\nu\rho} \, (1-\gamma_5) \,T^a_S\,,
\label{eq:ffWWg}
\end{equation}
\\

\vspace{.3cm}
\item \hspace{1cm}\\
\begin{picture}(55,45) (30,-30)
\SetWidth{0.5}
\ArrowLine(30,-30)(50,-10)
\ArrowLine(50,-10)(30,11)
\Gluon(70,-30)(50,-10){2}{5}
\Photon(70,11)(50,-10){2}{3}
\Gluon(75,-10)(50,-10){2}{5}
\Vertex(50,-10){1.5}
\Text(40,11)[lb]{$q$}
\Text(40,-40)[lb]{$q$}
\Text(80,-15)[lb]{$G^a_{\nu}(k_2)$}
\Text(77,-40)[lb]{$G^b_{\rho}(k_3)$}
\Text(77,8)[lb]{$A_{\mu}(k_1)$}
\end{picture}
\vspace{-1.5cm}
\begin{equation}
\hspace{2cm}
-\frac{i}{2}e\,g^2_s \,Q_q
\theta_{\mu\nu\rho} \,f^{abc}T^c_S\,,
\label{eq:ffAGG}
\end{equation}
\\

\vspace{.3cm}
\item \hspace{1cm}\\
\begin{picture}(55,45) (30,-30)
\SetWidth{0.5}
\ArrowLine(30,-30)(50,-10)
\ArrowLine(50,-10)(30,11)
\Gluon(70,-30)(50,-10){2}{5}
\Photon(70,11)(50,-10){2}{7}
\Gluon(75,-10)(50,-10){2}{5}
\Vertex(50,-10){1.5}
\Text(40,11)[lb]{$q$}
\Text(40,-40)[lb]{$q$}
\Text(80,-15)[lb]{$G^a_{\nu}(k_2)$}
\Text(77,-40)[lb]{$G^b_{\rho}(k_3)$}
\Text(77,8)[lb]{$Z_{\mu}(k_1)$}
\end{picture}
\vspace{-1.5cm}
\begin{equation}
\hspace{2cm}
-\frac{i\, e\,g^2_s }{2\sin2\theta_W}
\theta_{\mu\nu\rho} \, (c_{V,q}-c_{A,q}\gamma_5) \,f^{abc}T^c_S\,,
\label{eq:ffGGZ}
\end{equation}
\\

\vspace{.3cm}
\item\hspace{1cm}\\
\hspace*{1cm}
\begin{picture}(55,45) (30,-30)
\SetWidth{0.5}
\ArrowLine(30,-30)(50,-10)
\ArrowLine(50,-10)(30,11)
\Gluon(70,-30)(50,-10){2}{5}
\Photon(70,11)(50,-10){2}{7}
\Photon(75,-10)(50,-10){2}{3}
\Vertex(50,-10){1.5}
\Text(5,11)[lb]{${u}^{(i)}$}
\Text(3,-40)[lb]{${d}^{(j)}$}
\Text(80,-15)[lb]{$A_{\nu}(k_2)$}
\Text(77,-40)[lb]{$G^a_{\rho}(k_3)$}
\Text(77,8)[lb]{$W^+_{\mu}(k_1)$}
\end{picture}
\hspace*{4cm}
\begin{picture}(55,45) (30,-30)
\SetWidth{0.5}
\ArrowLine(30,-30)(50,-10)
\ArrowLine(50,-10)(30,11)
\Gluon(70,-30)(50,-10){2}{5}
\Photon(70,11)(50,-10){2}{7}
\Photon(75,-10)(50,-10){2}{3}
\Vertex(50,-10){1.5}
\Text(3,11)[lb]{${d}^{(j)}$}
\Text(5,-40)[lb]{${u}^{(i)}$}
\Text(77,-40)[lb]{$G^a_{\rho}(k_3)$}
\Text(80,-15)[lb]{$A_{\nu}(k_2)$}
\Text(77,8)[lb]{$W^-_{\mu}(k_1)$}
\end{picture}
\\

\hspace{-2.cm}
\begin{equation}
\hspace{-1cm}
\frac{e^2\,g_s }{4\sqrt2\sin\theta_W}\,
\left(
V_{ij} \atop
-V^{*}_{ij}
\right)\,
\theta_{\mu\nu\rho} \, (1-\gamma_5)\,
\,T^a_S\,,
\label{eq:ffAWg}
\end{equation}
\\

\vspace{.3cm}
\item \hspace{1cm}\\
\hspace*{1cm}
\begin{picture}(55,45) (30,-30)
\SetWidth{0.5}
\ArrowLine(30,-30)(50,-10)
\ArrowLine(50,-10)(30,11)
\Gluon(70,-30)(50,-10){2}{5}
\Photon(70,11)(50,-10){2}{7}
\Photon(75,-10)(50,-10){2}{7}
\Vertex(50,-10){1.5}
\Text(5,11)[lb]{${u}^{(i)}$}
\Text(3,-40)[lb]{${d}^{(j)}$}
\Text(80,-15)[lb]{$Z_{\nu}(k_2)$}
\Text(77,-40)[lb]{$G^a_{\rho}(k_3)$}
\Text(77,8)[lb]{$W^+_{\mu}(k_1)$}
\end{picture}
\hspace*{4cm}
\begin{picture}(55,45) (30,-30)
\SetWidth{0.5}
\ArrowLine(30,-30)(50,-10)
\ArrowLine(50,-10)(30,11)
\Photon(70,-30)(50,-10){2}{7}
\Gluon(70,11)(50,-10){2}{5}
\Photon(75,-10)(50,-10){2}{7}
\Vertex(50,-10){1.5}
\Text(3,11)[lb]{${d}^{(j)}$}
\Text(5,-40)[lb]{${u}^{(i)}$}
\Text(77,-40)[lb]{$G^a_{\rho}(k_3)$}
\Text(80,-15)[lb]{$Z_{\nu}(k_2)$}
\Text(77,8)[lb]{$W^-_{\mu}(k_1)$}
\end{picture}

\hspace{-1.5cm}
\begin{eqnarray}
\hspace{-1.5cm}
\frac{e^2\,g_s\,\cos\theta_W}{4\sqrt2 \sin^2\theta_W}\,
\left(
V_{ij} \atop
-V^{*}_{ij}
\right)\,
\theta_{\mu\nu\rho} \, (1-\gamma_5)
\, T^a_S\,,
\nonumber
\\
\label{eq:ffGWZ}
\end{eqnarray}
\\
\item \hspace{1cm}\\
\hspace*{1cm}
\begin{picture}(55,45) (30,-30)
\SetWidth{0.5}
\ArrowLine(30,-30)(50,-10)
\ArrowLine(50,-10)(30,11)
\Gluon(70,-30)(50,-10){2}{5}
\Photon(70,11)(50,-10){2}{7}
\Gluon(75,-10)(50,-10){2}{5}
\Vertex(50,-10){1.5}
\Text(5,11)[lb]{${u}^{(i)}$}
\Text(3,-40)[lb]{${d}^{(j)}$}
\Text(80,-15)[lb]{$G^a_{\nu}(k_2)$}
\Text(77,-40)[lb]{$G^b_{\rho}(k_3)$}
\Text(77,8)[lb]{$W^+_{\mu}(k_1)$}
\end{picture}
\hspace*{4cm}
\begin{picture}(55,45) (30,-30)
\SetWidth{0.5}
\ArrowLine(30,-30)(50,-10)
\ArrowLine(50,-10)(30,11)
\Gluon(70,-30)(50,-10){2}{5}
\Photon(70,11)(50,-10){2}{7}
\Gluon(75,-10)(50,-10){2}{5}
\Vertex(50,-10){1.5}
\Text(3,11)[lb]{${d}^{(j)}$}
\Text(5,-40)[lb]{${u}^{(i)}$}
\Text(80,-15)[lb]{$G^a_{\nu}(k_2)$}
\Text(77,-40)[lb]{$G^b_{\rho}(k_3)$}
\Text(77,8)[lb]{$W^-_{\mu}(k_1)$}
\end{picture}

\hspace{-1.5cm}
\begin{equation}
\hspace{-2cm}
\frac{-i\,e\,g^2_s }{4\sqrt2\sin\theta_W}
\left(
V_{ij}\atop
V^{*}_{ij}
\right)\,
\theta_{\mu\nu\rho} \, (1-\gamma_5)
\,
\, f^{abc}\,T^c_S\,.
\label{eq:ffGGW}
\end{equation}
\\
\end{itemize}

The Feynman rules for pure QCD vertices, (\ref{eq:ffgluon})-(\ref{eq:ffGG}) 
have already been given in \cite{Carlson:2001sw}. 
\subsection{\label{sec:FRnon-minimal}Non-Minimal NCSM}

Here we give the Feynman rules for
$\theta$-corrections to gauge boson vertices
$Zgg$ and $\gamma gg$
in the non-minimal NCSM introduced in \cite{Behr:2002wx}.
Observe that the quark sector is not affected by the change of
the representation in the gauge part of the action, and,
consequently, is the same in both models, the mNCSM and the nmNCSM.
\noindent
\begin{itemize}
\item\hspace{1cm}\\
\begin{picture}(55,45) (30,-30)
\SetWidth{0.5}
\Photon(30,-30)(50,-10){2}{3}
\Gluon(50,-10)(30,11){2}{5}
\Gluon(80,-10)(50,-10){2}{5}
\Vertex(50,-10){1.5}
\Text(40,11)[lb]{$G^b_{\rho}(k_3)$}
\Text(40,-40)[lb]{$A_{\mu}(k_1)$}
\Text(75,-5)[lb]{$G^a_{\nu}(k_2)$}
\end{picture}
\\
\vspace{-1.5cm}
\begin{eqnarray}
\hspace{2cm}
-2\,e\,\sin{2\theta_W}{\rm K}_{\gamma gg}  \,
\Theta_3((\mu,k_1),(\nu,k_2),(\rho,k_3))\,\delta^{ab}\,,
\label{nm2Wgamma}
\end{eqnarray}
\vspace{.5cm}
\item\hspace{1cm}\\

\begin{picture}(55,45) (30,-30)
\SetWidth{0.5}
\Photon(30,-30)(50,-10){2}{7}
\Gluon(50,-10)(30,11){2}{5}
\Gluon(80,-10)(50,-10){2}{5}
\Vertex(50,-10){1.5}
\Text(40,11)[lb]{$G^b_{\rho}(k_3)$}
\Text(40,-40)[lb]{$Z_{\mu}(k_1)$}
\Text(75,-5)[lb]{$G^a_{\nu}(k_2)$}
\end{picture}
\\
\vspace{-1.5cm}
\begin{eqnarray}
\hspace{2cm}
- 2\,e\,\sin{2\theta_W}{\rm K}_{Zgg}  \,
\Theta_3((\mu,k_1),(\nu,k_2),(\rho,k_3))\,\delta^{ab}\,.
\label{nm2WZ}
\end{eqnarray}


\end{itemize}

The constants K are not independent,
and they are defined in Eqs. \req{K123456},
from where
${\rm K}_{Zgg}=-\tan{\theta_W}{\rm K}_{\gamma gg}$.

\section{\label{sec:con}Conclusions}
This article, together with \cite{Melic:2005fm}, represents a complete 
description of the Non-Commutative Standard Model constructed in
\cite{Calmet:2001na,Behr:2002wx} and makes it accessible to further research. 

We have presented a
careful discussion of QCD-electroweak charged and neutral
currents as well as a detailed analysis of the Yukawa part of the
NCSM action. The NCSM action is expressed in
terms of physical fields and mass eigenstates,
up to the first order in the non-commutative parameter $\theta$.

The novel feature of the NCSM presented in this article
is the presence of mixtures of gluon
interactions with electroweak ones. 
We again encounter the mass-proportional corrections to 
the quark-boson couplings, stemming from the Yukawa part of the action (\ref{eq:SYukawa-fiz}). 
These features are introduced by the Seiberg-Witten maps.

With these interactions provided, together with \cite{Melic:2005fm}, we complete 
the analysis of all interactions of the NCSM appearing at the order $\theta$, and 
hope that this will enable further investigations leading to the stringent 
bounds to the non-commutativity parameters.

\subsection*{Acknowledgment}

We want to thank Julius Wess for many fruitful discussions. 
M.W. also
wants to acknowledge the support from ``Fonds zur F\"orderung der
wissenschaftlichen Forschung" (Austrian Science Fund), projects
P15463-N08 and P16779-N02. This work was partially supported by the
Ministry of Science, Education and Sport of the Republic of Croatia.


\begin{thebibliography}{10}


\bibitem{Melic:2005fm}
B.~Meli\'c, K.~Passek-Kumeri\v{c}ki, J.~Trampeti\'c, P.~Schupp 
and M.~Wohlgenannt,
hep-ph/0502249.

\bibitem{Calmet:2001na}
X.~Calmet, B.~Jur\v{c}o, P.~Schupp, J.~Wess and M.~Wohlgenannt, 
Eur.~Phys.~J. C{\bf 23} (2002) 363
[hep-ph/0111115].

\bibitem{Kontsevich:1997vb}
M.~Kontsevich,
Lett.\ Math.\ Phys.\  {\bf 66} (2003) 157
[q-alg/9709040].

\bibitem{Seiberg:1999vs}
N. Seiberg and E. Witten, 
JHEP {\bf 09} (1999) 032 [hep-th/9908142].

\bibitem{Madore:2000en}
J. Madore, S. Schraml, P.~Schupp and J.~Wess, 
Eur. Phys. J. C{\bf 16} (2000) 161 [hep-th/0001203].

\bibitem{Jurco:2000ja}
B. Jur\v{c}o, S. Schraml, P.~Schupp and J.~Wess, 
Eur. Phys. J. C{\bf 17} (2000) 521 [hep-th/0006246].

\bibitem{Jurco:2001rq}
B. Jur\v{c}o, L. M\"oller, S.~Schraml, P.~Schupp and J.~Wess, 
Eur. Phys. J. C{\bf 21} (2001) 383 [hep-th/0104153].

\bibitem{Aschieri:2002mc}
P. Aschieri, B.~Jur\v{c}o, P.~Schupp and J.~Wess, 
Nucl. Phys. B{\bf 651} (2003) 45 [hep-th/0205214].

\bibitem{Chaichian:2001aa}
M.~Chaichian, P.~Presnajder, M.M.~Sheikh-Jabbari and A.~Tureanu, 
Eur.~Phys. J. C{\bf 29} (2003) 413 [hep-th/0107055].

\bibitem{Chaichian:2004za}
M.~Chaichian, P.~P.~Kulish, K.~Nishijima and A.~Tureanu,
Phys. Lett. B{\bf 604} (2004) 98 [hep-th/0408069].

\bibitem{Chaichian:2004yw}
M.~Chaichian, A.~Kobakhidze and A.~Tureanu,
hep-th/0408065.

\bibitem{Mocioiu:2000ip}
I.~Mocioiu, M.~Pospelov and R.~Roiban,
Phys.\ Lett.\ B {\bf 489} (2000) 390
[hep-ph/0005191].

\bibitem{Carlson:2001sw}
C.~E.~Carlson, C.~D.~Carone and R.~F.~Lebed,
Phys.\ Lett.\ B {\bf 518} (2001) 201
[hep-ph/0107291].

\bibitem{Anisimov:2001zc}
A.~Anisimov, T.~Banks, M.~Dine and M.~Graesser,
Phys.\ Rev.\ D {\bf 65} (2002) 085032
[hep-ph/0106356].

\bibitem{Armoni:2000xr} 
A.~Armoni, 
Nucl. Phys. B{\bf 593} (2001) 229
[hep-th/0005208].

\bibitem{Behr:2002wx}
W.~Behr, N.G. Deshpande, G. ~Duplan\v{c}i\'{c}, P.~Schupp, J.~Trampeti\'{c} and
J.~Wess, 
Eur. Phys. J. C{\bf 29} (2003) 441 [hep-ph/0202121].

\bibitem{Duplancic:2003hg}
G.~Duplan\v{c}i\'{c}, P.~Schupp and J.~Trampeti\'{c}, 
Eur.~Phys. J. C{\bf 32} (2003) 141 [hep-ph/0309138].

\bibitem{Schupp:2002up}
P.~Schupp, J.~Trampeti\'{c}, J.~Wess and G.~Raffelt,
Eur.\ Phys.\ J.\ C {\bf 36} (2004) 405
[hep-ph/0212292].

\bibitem{Minkowski:2003jg}
P.~Minkowski, P.~Schupp and J.~Trampeti\'{c},
Eur.\ Phys.\ J.\ C {\bf 37} (2004) 123
[hep-th/0302175].

\bibitem{Trampetic:2002eb}
J. Trampeti\'c, 
Acta Phys. Polon. B{\bf 33}
(2002) 4317 [hep-ph/0212309].

\bibitem{Schupp:2004dz}
P. Schupp and J. Trampeti\'c, 
hep-ph/0405163.

\bibitem{Ohl:2004tn}
T.~Ohl and J.~Reuter,
Phys. Rev. D{\bf 70} (2004) 076007 [hep-ph/0406098].

%
%
%
%
%

%
%
%
%
%
%
%
\bibitem{Brandt:2003fx}
F. Brandt, C.P. Mart\'{i}n and F. Ruiz Ruiz,
JHEP {\bf 07} (2003) 068
[hep-th/0307292].

%
\end{thebibliography}
\end{document}